\documentclass[twocolumn]{aa}
\usepackage[pdftex]{xcolor}
\usepackage{graphics,graphicx}
\usepackage{multirow}
\usepackage{eurosym}
\usepackage{enumitem}
\usepackage{natbib}      
\usepackage{dashrule}    

\def\asec{\ifmmode ^{\prime\prime}\else$^{\prime\prime}$\fi}
\def\amin{\ifmmode ^{\prime}\else$^{\prime}$\fi}

\def\lsim{\mathrel{\rlap{\lower4pt\hbox{\hskip1pt$\sim$}}
    \raise1pt\hbox{$<$}}}                
\def\gsim{\mathrel{\rlap{\lower4pt\hbox{\hskip1pt$\sim$}}
    \raise1pt\hbox{$>$}}}                

\def\degs{\ifmmode ^{\circ}\else$^{\circ}$\fi}

\newcommand{\gp}{\mbox{$g^\prime$}}
\newcommand{\rp}{\mbox{$r^\prime$}}
\newcommand{\ip}{\mbox{$i^\prime$}}
\newcommand{\zp}{\mbox{$z^\prime$}}

\begin{document}

\title{GRB 151027B - large-amplitude late-time radio variability\thanks{This paper makes use of the following data: ATCA: Proposal C2955 (PI: Greiner), 
ALMA: ADS/JAO.ALMA\#2015.1.01558.T (PI: Schulze).}}

\titlerunning{GRB 151027B}

\author{J. Greiner\inst{1}, J. Bolmer\inst{1,2}, M. Wieringa\inst{3},
A.J. van der Horst\inst{4,5}, D. Petry\inst{6}, S. Schulze\inst{7}, 
F. Knust\inst{1}, 
G. de Bruyn\inst{8}\thanks{deceased}, T. Kr\"uhler\inst{1}, 
P. Wiseman\inst{1}\thanks{Present address: School of Physics and Astronomy, Univ. of Southampton, Southampton, S017 1BJ, UK}, S. Klose\inst{9},
C. Delvaux\inst{1}, J.F. Graham\inst{1}, D.A. Kann\inst{9, 10}, 
A. Moin\inst{11}, A. Nicuesa-Guelbenzu\inst{9},
P. Schady\inst{1}, S. Schmidl\inst{9}, T. Schweyer\inst{1},
M. Tanga\inst{1}, S. Tingay\inst{12}, H. van Eerten\inst{13},
K. Varela\inst{1}
 }

\authorrunning{Greiner et al.}

\offprints{J. Greiner, jcg@mpe.mpg.de}

\institute{
  Max-Planck-Institut f\"ur extraterrestrische Physik, 
    Giessenbachstr. 1, 85748 Garching, Germany
  \and
  European Southern Observatory, Alonso de Cordova 3107, Vitacura, Casilla 
  19001, Santiago 19, Chile
  \and
  CSIRO Astronomy \& Space Science, Locked Bag 194, Narrabri,
             NSW 2390, Australia
  \and 
  Dept. of Physics, The George Washington University, Staughton Hall, 
   707 22nd Street NW, Washington, D.C. 20052, U.S.A.
  \and
   Astronomy, Physics, and Statistics Institute of Sciences (APSIS),
   Staughton Hall, 
   707 22nd Street NW, Washington, D.C. 20052, U.S.A.
  \and
  ALMA Regional Centre, European Southern Observatory, 
  Karl-Schwarzschild-Strasse 2, 85748 Garching, Germany 
  \and
  Dept. of Particle Physics and Astrophysics, Faculty of Physics,
  Weizmann Institute of Science, Rehovot 76100, Israel
  \and
  ASTRON, Postbus 2, 7900 AA Dwingeloo, The Netherlands
  \and
  Th\"uringer Landessternwarte Tautenburg, Sternwarte 5, 07778 Tautenburg, 
    Germany
  \and
    Instituto de Astrof\'isica de Andaluc\'ia (IAA-CSIC), Glorieta de la 
    Astronomia s/n, 18008 Granada, Spain
  \and
  Physics Department, United Arab Emirates University, P.O. Box 15551,
  Al-Ain, United Arab Emirates
  \and
  Curtin Institute of Radio Astronomy, GPO Box U1987, Perth,
             Western Australia, 6845, Australia
  \and
  Dept. of Physics, University of Bath, Claverton Down, Bath BA2 7AY, 
  United Kingdom
}

\date{Received 10 August 2017 / Accepted 06 February 2018 }

\abstract{Deriving physical parameters from gamma-ray burst afterglow
observations remains a challenge, even now, 20 years after the
discovery of afterglows. The main reason for the lack of progress is
that the peak of the synchrotron emission is in the sub-mm range, thus
requiring radio observations in conjunction with X-ray/optical/near-infrared
data in order to measure the corresponding spectral slopes and 
consequently remove the ambiguity wrt. slow vs. fast cooling and the
ordering of the characteristic frequencies.
}
{We have embarked on a multi-frequency, multi-epoch observing campaign 
to obtain sufficient data for a given GRB  that allows us to
test the simplest version of the fireball afterglow model.
}
{We observed GRB 151027B, the 1000th \textit{Swift}-detected GRB, with GROND
in the optical-NIR, ALMA in the sub-millimeter, ATCA in the radio band,
and combine this with public \textit{Swift}-XRT X-ray data.
}
{While some observations at crucial times only return upper limits
or surprising features,
the fireball model is narrowly constrained by our data set, and allows
us to draw a consistent picture with a fully-determined parameter set.
Surprisingly, we find rapid, large-amplitude flux density variations
in the radio band which are extreme not only for GRBs, but generally
for any radio source. We interpret these as scintillation effects,
though the extreme nature requires either the scattering screen to be at
much smaller distance than usually assumed, multiple screens, or a 
combination of the two.
}
{The data are consistent with the simplest fireball scenario, for
a blast wave moving into a constant-density medium, and slow-cooling.
All fireball parameters are constrained to better than or about a factor 
of two, except for the density and the fraction of the energy in the 
magnetic field which has a factor 10 uncertainty in both directions. 
}


\maketitle

\section{Introduction}

Long-duration Gamma-Ray Bursts (GRBs) are widely accepted to be 
related to the death of massive stars \citep{Hjorth03, Stanek03}.
Due to their large $\gamma$-ray luminosity they can be
detected to very high redshift, and 
thus provide a unique probe into the early Universe.
How the afterglow emission evolves both in frequency space and with time
depends on both the properties of the burst environment
(e.g., gas density profile, dust)
and the progenitor itself (e.g., temporal energy injection profile as well
as mass, rotation, and binarity, all of which influence the 
density and structure of the circumburst medium, e.g., \citealt{Yoon+2012}).

When the relativistically expanding blast wave interacts with the 
circum-burst medium, an external shock is formed,
the macroscopic properties of which are well understood.
Under the implicit assumptions that the electrons are ``Fermi'' accelerated 
at the relativistic shock to a power-law distribution, 
their dynamics can be expressed
in terms of four main parameters:
(1) the total internal energy in the shocked region as released in the 
explosion, 
(2) the electron density $n$ and radial profile of the  
surrounding medium, 
(3) the fraction of the shock energy that goes into electrons $\epsilon_e$,
(4) the ratio of the magnetic field energy density to the total energy,
 $\epsilon_B$.
Measuring the energetics or the energy partition ($\epsilon_e$/$\epsilon_B$)
has been challenging, and observations at multiple different passbands have
thus far only been possible for a dozen 
of the more than 1000 GRB afterglows detected so far.

The observational difficulty of
establishing whether the observed synchrotron spectrum is in the fast or
slow cooling stage introduces a degeneracy when attempting to explain 
the spectrum in terms of the physical model parameters.
The minimal and simplest afterglow model has five parameters 
(not counting the distance/redshift).
The degeneracy between many of these parameters makes it even more
difficult to draw firm conclusions.
Thus, it is not surprising that many previous
attempts had to compromise whenever assumptions had to be made 
about individual parameters 
\citep[e.g.,][]{PanaiKumar02, Yost03, Chandra2008, Cenko10, 
gkn13, Laskar14, Varela+2016},
but contradictions between analyses with different assumptions
surfaced only in the rare cases when the same 
GRB afterglows were analyzed based on different data sets
\citep[e.g.,][]{mkr10, Cenko11}. 

Here, we report our multi-epoch, multi-frequency observations of
GRB 151027B, in an attempt to collect an exhaustive dataset which
would allow us to determine all these parameters. 

\section{GRB 151027B detection and afterglow observations}

\subsection{GRB prompt and afterglow detection}

GRB 151027B was detected by the \textit{Swift} \citep{gcg04} 
Burst Alert Telescope (BAT, \citealt{Barthelmy05}) on 2015 October 27 at 
$T_0$ = 22:40:40 UT (MJD = 57322.944907) as the 1000th 
\textit{Swift} burst \citep{Ukwatta2015}. The prompt light curve shows 
a complex structure with several overlapping peaks that starts at $\sim T_0$
and extends for about 100 s, leading to a formal duration T90 
(15--350 keV) of 80$\pm$36 s
\citep{Sakamoto2015}. \textit{Swift} slewed immediately to the
BAT-derived position, allowing the X-ray afterglow to be discovered
readily with the \textit{Swift} X-ray telescope (XRT, \citet{bur05}) 
with a 4\asec\ accurate position (later refined to 1\farcs8). 
This in turn allowed the
discovery of the optical afterglow one hour later by the Nordic Optical
Telescope \citep{Malesani2015}, and a redshift determination of $z = 4.063$
with VLT/X-shooter another four hours later \citep{Xu2015}. In addition to our 
GROND observations (see below), detections of the optical afterglow
were also reported by MASTER \citep{Buckley2015}, RATIR \citep{Watson2015}
and the 2-m Faulkes Telescope North in Hawaii \citep{Dichiara2015}.
\textit{Swift}-UVOT did not detect the afterglow, consistent with 
the redshift and galactic foreground extinction \citep{Breeveld2015}.

\begin{figure}[th]
\includegraphics[width=1.0\columnwidth]{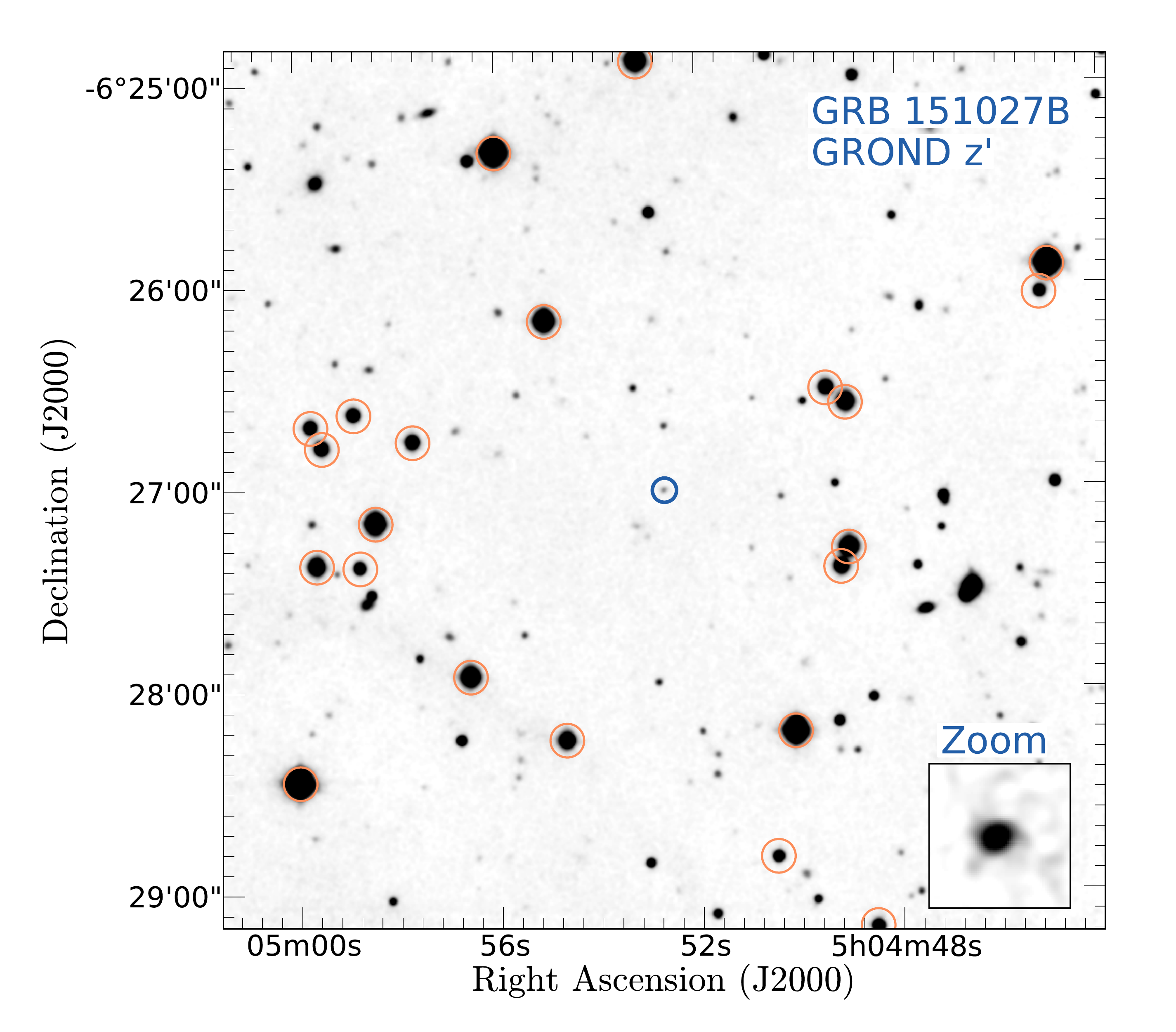}
\caption[151027Bfc]{Finding chart of the afterglow of GRB 151027B 
based on a GROND \zp-band image, with secondary standard stars
of Tab. \ref{Stdstars} encircled.
North is up, and East to the left.
\label{fc}}
\end{figure}

\begin{table*}[th]
   \caption{GROND observations; all in the AB system, not corrected for 
     Galactic foreground extinction corresponding to $E(B-V) = 0.18$ mag
     ($A_V = 0.55$ mag) \citep{SchlaflyFinkbeiner2011}.}
   \vspace{-0.2cm}
      \begin{tabular}{cccccccc}
      \hline
      \noalign{\smallskip}
      Time after $T_0$ & \gp & \rp & \ip & \zp &   $J$ &  $H$  & $K_{\rm s}$ \\ 
      (s)              &(mag)&(mag)&(mag)&(mag)&(mag)&(mag)&(mag)  \\
      \noalign{\smallskip}
      \hline
      \noalign{\smallskip}
$28830\pm 903$ 		& $22.16\pm 0.23$	& $20.41\pm 0.05$ 	& $19.78\pm 0.04$ 	& $19.58\pm 0.04$ 	& $19.47\pm 0.09$ 	& $18.96\pm 0.08$ 	& $19.09\pm 0.16$ 	\\
$30948\pm 1125$ 	& $22.17\pm 0.26$ 	& $20.59\pm 0.07$ 	& $20.02\pm 0.06$ 	& $19.66\pm 0.05$	& $19.48\pm 0.10$ 	& $19.19\pm 0.10$ 	& $19.10\pm 0.20$  	\\
$33291\pm 1129$ 	& $22.69\pm 0.24$ 	& $20.62\pm 0.04$ 	& $20.00\pm 0.04$ 	& $19.67\pm 0.04$ 	& $19.52\pm 0.07$ 	& $19.17\pm 0.08$ 	& $19.10\pm 0.14$	\\
$35634\pm 1130$ 	& $22.45\pm 0.16$ 	& $20.64\pm 0.04$ 	& $20.08\pm 0.04$ 	& $19.77\pm 0.04$ 	& $19.54\pm 0.07$ 	& $19.39\pm 0.08$ 	& $19.24\pm 0.14$ 	\\
$112431\pm 6317$ 	& $>23.6$ 			& $22.05\pm 0.07$  	& $21.53\pm 0.07$ 	& $21.18\pm 0.06$ 	& $21.10\pm 0.15$ 	& $20.73\pm 0.16$ 	& $20.36\pm 0.24$ 	\\
$202273\pm 1606$ 	& $>24.0$ 			& $22.57\pm 0.09$ 	& $22.09\pm 0.09$ 	& $21.85\pm 0.08$ 	& $21.80\pm 0.25$ 	& $21.53\pm 0.29$ 	& $>20.9$		\\ 
$804300\pm 6436$ 	& $>25.8$ 			& $24.75\pm 0.16$ 	& $24.33\pm 0.21$ 	& $23.87\pm 0.23$ 	& $>22.3$ 			& $>21.9$ 			& $>21.5$ \\ 
$\!\!1838665\pm 3570$   & $>25.5$               & $>25.7$ 	& $>24.9$ 	& $>24.7$ 	& $>22.4$ 			& $>22.0$       & $>21.4$ \\
     \noalign{\smallskip}
      \hline
   \end{tabular}
   \label{GROND}
\end{table*}

\subsection{GROND observations}

Observations with GROND \citep{gbc08} started on 2015 October 28 at 06:26 UT, 
about 8 hr after the trigger, at a Moon distance of only 37\degs. Simultaneous 
imaging in $g'r'i'z'JHK_{\rm s}$ continued for several further epochs 
(see the observation log in Tab. \ref{GROND}) until 
2015 November 18, when the afterglow could not be detected anymore.
During the night of November 5/6, a field with SDSS coverage 
(RA(2000.0)=03h 45m, Decl.(2000.0)=-06\degs15\amin) was observed
immediately after the GRB field under photometric conditions.

\begin{table*}[t]
  \caption{ATCA observing details}
  \vspace{-0.25cm}
  \begin{tabular}{lccccccc}
    \hline
    \noalign{\smallskip}
    Date \& Start-Time & On source      & Time after GRB & Telescope & 5.5 GHz flux & 9 GHz flux \\
                & exposure (hr) &  (days)$^{a)}$  & configuration& $\mu$Jy$^{b)}$ & $\mu$Jy$^{b)}$ \\
    \noalign{\smallskip}
    \hline
    \noalign{\smallskip}
2015-10-30 12:56 & 3.3 & \,\,2.76$\pm$0.16  & 6A   & $<$18     &  67$\pm$10 \\
2015-11-02 12:00 & 3.2 & \,\,5.74$\pm$0.19  & 6A   & 73$\pm$10 &  98$\pm$11 \\
2015-11-11 15:36 & 5.8 & 14.70$\pm$0.19  & 6A   & 76$\pm$7  &  $<$15     \\
2015-11-14 11:31 & 3.3 & 17.69$\pm$0.16  & 6A   & $<$26.0   & 100$\pm$10 \\
2015-11-16 12:00 & 6.7 & 19.72$\pm$0.17  & 1.5A & $<$13.4   & $<$15.4   \\
2015-12-02 09:36 & 8.4 & 35.67$\pm$0.21  & 1.5A & 60$\pm$11 & 36$\pm$11 \\
2015-12-11 10:05 & 2.5 & 44.54$\pm$0.06  & 750C & 71$\pm$12 & $<$22   \\
2016-01-22 06:30 & 5.7 & 86.48$\pm$0.16  & EW352& $<$24     & $<$28   \\
   \noalign{\smallskip}
     \hline
  \end{tabular}

  $^{a)}$ The ``error'' denotes the time span over which the exposure was
     spread to cover the $u-v$ plane.
  $^{b)}$ Upper limits are given at the 2$\sigma$ level.
  \label{radio}
\end{table*}

GROND data have been reduced in the standard manner \citep{kkg08}
using pyraf/IRAF \citep{Tody1993, kkg08b}.
The optical/NIR imaging was calibrated against the primary 
SDSS\footnote{http://www.sdss.org} 
standard star network, or catalogued magnitudes of field stars from the 
SDSS in the case of $g^\prime r^\prime i^\prime z^\prime$ observations,
or the 2MASS catalog for $JHK_{\rm s}$ imaging. This results in typical 
absolute accuracies  of $\pm$0.03~mag in $g^\prime r^\prime i^\prime 
z^\prime$ and $\pm$0.05~mag in $JHK_{\rm s}$. 
Comparison stars covered by the finding chart of GRB 151027B (Fig. \ref{fc})
are given in Tab. \ref{Stdstars}.

Despite its high redshift, the afterglow was detected in all seven bands 
(Tab. \ref{GROND})
at a common position of RA(2000.0), Dec(2000.0) = 76\fdg21955, -6\fdg45029, 
or 05:04:52.69 --06:27:01.1, with a 1$\sigma$ error of $\pm$0\farcs25.
This is fully consistent with both, the UVOT-corrected Swift/XRT position
\citep{Osborne2015} as well as the NOT-derived position \citep{Malesani2015}.

\subsection{ATCA observations}

We observed the field of GRB 151027B under program C2955 
(PI: Greiner) simultaneously at 5.5 and 9 GHz 
with the Australia Telescope Compact Array (ATCA), beginning at 
October 30.54 UT for 3.3 hr; the corresponding detection at 9 GHz
has been reported earlier \citep{Greiner2015}. Over the following three months,
we observed the GRB 151027B position at another 7 epochs. A summary
of the observing log, including the telescope configuration, 
is given in Tab. \ref{radio}.
The observations were mostly performed with the CFB 1M-0.5K mode,
providing 2048 channels per 2048 MHz continuum intermediate frequency (IF;
1 MHz resolution) and 2048 channels per 1 MHz zoom band 
(0.5 kHz resolution).
Data analysis was done using the standard software package \texttt{MIRIAD}
\citep{stw95}, applying appropriate bandpass, phase and flux calibrations.
The quasar 0458-020 was used as phase and 1934-638 as flux calibrator.
Multifrequency synthesis images were constructed using robust weighting
(robust=0) and the full bandwidth between its flagged edges. 
The noise was determined by estimating the root-mean-square (rms) in
emission-free parts of the cleaned map.

\begin{figure}[th]
\hspace{-0.9cm}
\includegraphics[width=0.85\columnwidth, angle=270]{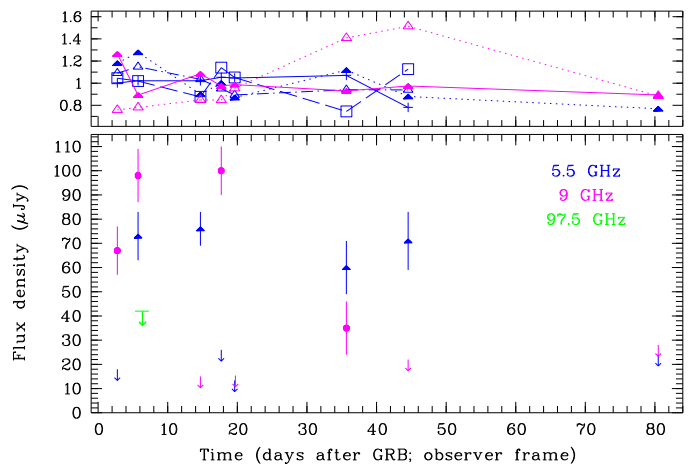}
\vspace{-0.9cm}
\caption[151027Bfc]{Radio light curve of the afterglow of GRB 151027B 
at 5.5 and 9 GHz, with the ALMA 97.5 GHz 2$\sigma$ upper limit overplotted
(lower panel). The upper panel shows the measured fluxes of selected 
brighter (130-500 $\mu$Jy) sources. While their nature or intrinsic
variability is not known, their $<$20\% flux variation demonstrates that
the strong fluctuations seen for the GRB 151027B afterglow (which would
correspond to an amplitude between 0.2 and 2 in this graph) is not 
an instrumental artifact.
\label{radiolc}}
\end{figure}

Given the largely varying flux levels between different observations
and also large flux differences between the two frequencies, we note
that the signal-to-noise ratio in most detections is so high that it is
unlikely our measurements are wrong. We employed 
two further tests:
1) for the November 14 observation, we made separate images for the top and
bottom of the 9 GHz band, resulting in flux measurements of 
96 $\mu$Jy (8-9 GHz) and 106 $\mu$Jy (9-10 GHz), thus providing an
internally consistent result;
2) we checked for other sources in the field for evidence of such 
 variation, but did not find any (see Fig. \ref{radiolc}).

\subsection{ALMA observations}

\begin{figure*}[th]
\hspace{-0.5cm}
\includegraphics[width=1.15\columnwidth, angle=0]{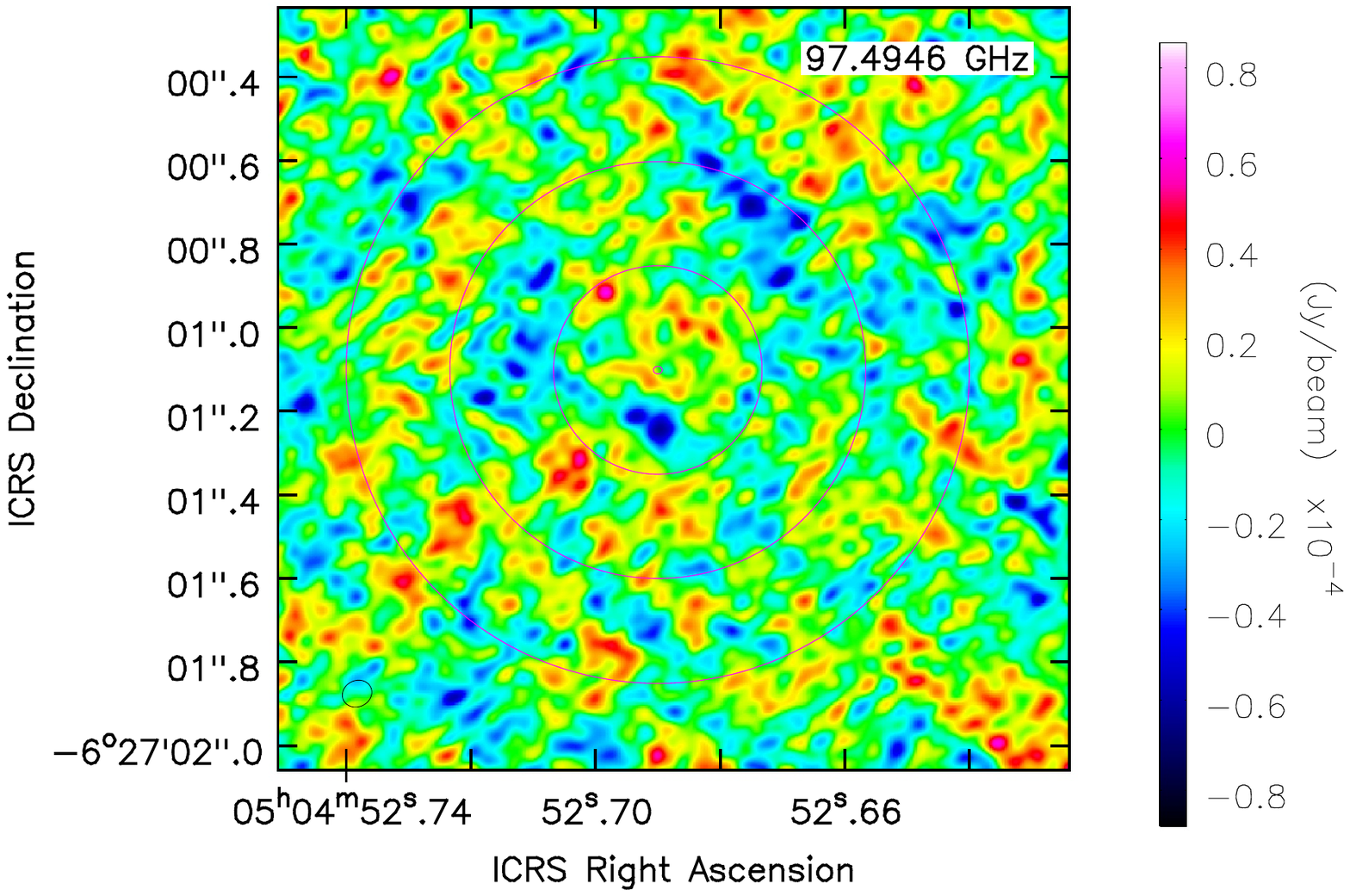}
\hspace{-1.2cm}
\includegraphics[width=1.15\columnwidth, angle=0]{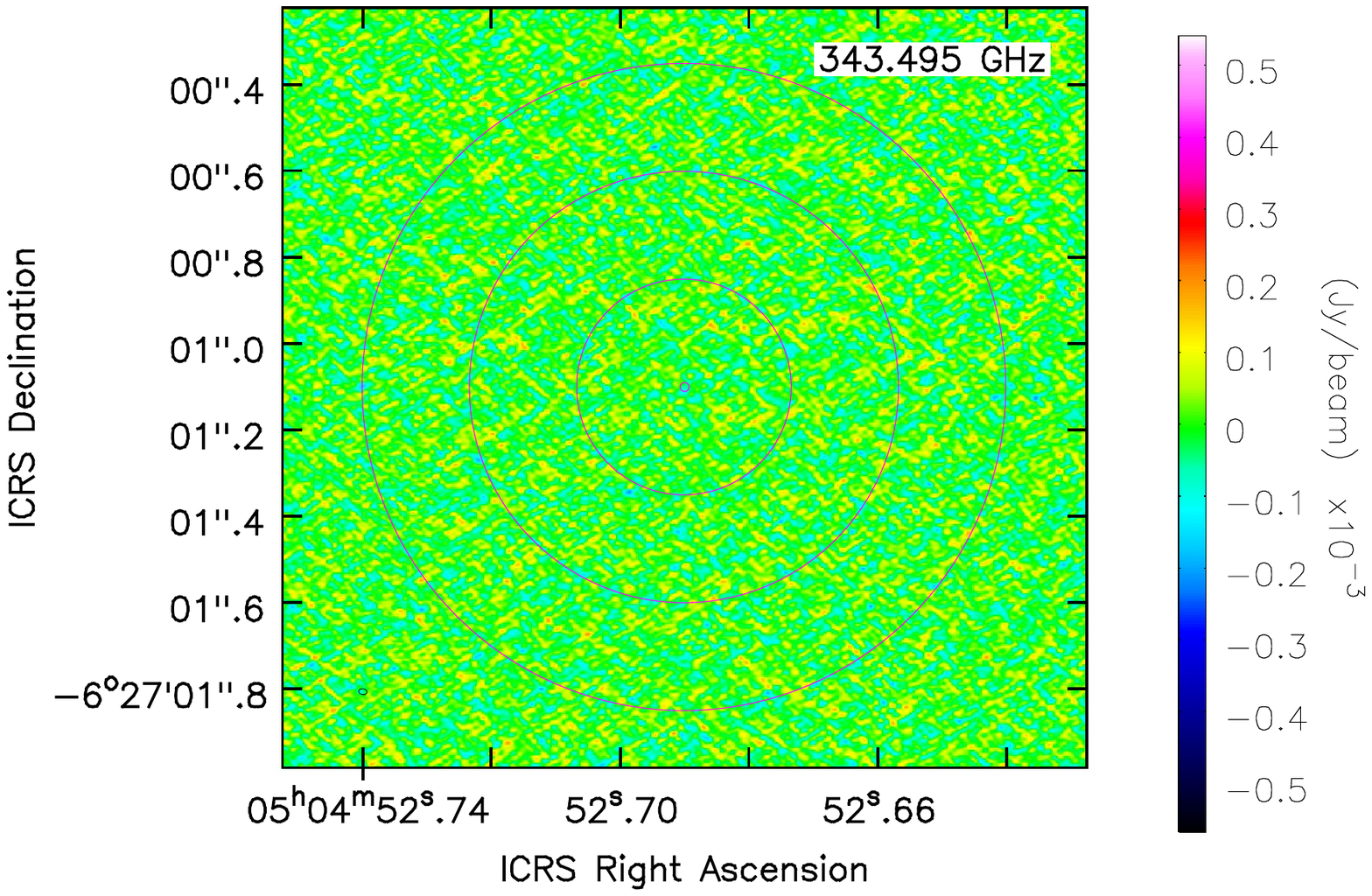}
\vspace*{-7.9cm}
\caption[151027BALMA]{ALMA images of the GRB 151027B location at band 3 
(97.5 GHz; left) and band 7 (343.5 GHz; right), in the International
Celestial Reference System (ICRS). The concentric circles 
around our best-fit GROND position are the 1$\sigma$, 2$\sigma$ and 3$\sigma$
error circles. The contour in the lower left of each figure gives the size
of the synthesized beam of the observation. It is due to the smaller beam
size that the band 7 image visually looks smoother than that of band 3; the
rms-noise is actually worse, as given in the text.
\label{almafig}}
\end{figure*}

ALMA observations were triggered under proposal-ID 2015.1.01558.T (PI: S. Schulze).
A band 7 (343.495 GHz) observation was performed starting on 
2015 November 2, 05:22 UT under a
precipitable water vapour (PWV) of 0.71 mm, and a band 3 (97.495 GHz)
observation 
started on 2015 November 4 at 07:45 UT under a PWV=0.31 mm.
The data analysis was performed using the standard ALMA data analysis
package CASA \citep{McMullin+2007, Petry+2012}, following the default
calibration path also used in ALMA Quality Assurance.
The final images are shown in Fig. \ref{almafig}.

Within the GROND error circle of 0\farcs25, in band 3, we find a peak with
a flux of 0.0619 mJy which given the rms noise of 0.0210 mJy corresponds
to nearly 3$\sigma$. However, the area of the error circle contains 
1914 spatial resolution pixels, so we expect 
1914 pixels $\times$ 0.0016 = 3 pixels to be above a 3$\sigma$ flux level.
Thus, the presence of the source-like point
in the error circle is compatible with a random occurrence, likely thermal 
noise. In fact, there are similar peaks outside the error circle.

In band 7, we find a peak within the error circle of 0.177 mJy
which given the rms noise of 0.0496 mJy corresponds
to 3.5$\sigma$. However, the area of the error circle contains 
7793 spatial resolution pixels, so we expect 
7793 pixels $\times$ 0.0002 = 1.6 pixels to be above a 
3.5$\sigma$ flux level.

Summarizing, no source is detected in either observation, 
with 2$\sigma$ upper limits of
42 $\mu$Jy in band 3 (97.495 GHz, integrated over a bandwidth of 7 GHz; 
taking into account that only about 87\% of each of the four spectral windows 
was used; edge channels are not good) at 7.378 days after the GRB,
and 100 $\mu$Jy in band 7 (343.495 GHz, also with a bandwidth of 7 GHz)
at 5.279 days after the GRB.
These values include the primary beam correction (though this is
$>$0.99 due to being close to the center of the field of view).

\section{Results}

Here, we will analyze our data in the context of the GRB fireball model
\citep{mer97, grs02}.
Throughout this paper, we use the definition 
$F_{\nu} \propto t^{-\alpha} \nu^{-\beta}$ where $\alpha$ is the 
temporal decay index, and $\beta$ is the spectral slope.

\subsection{Radio scintillation}

The large-amplitude radio variability observed in this GRB is very unusual. 
In the context of the canonical fireball scenario one would expect a 
smoothly varying afterglow, perhaps with a rapid rise and decay due to 
reverse shock emission, none of which is akin to our data. Moreover, 
we observe large variations between the simultaneously covered 5.5 and 
9~GHz bands, i.e. the inferred spectral slope changes between $<-2.3$ and 
$>2.9$ within days, while temporal slopes in the range $<-15$ and $>9$ over 
$2-3$ days are implied. We are not aware of any physical process(es) in 
GRB jets or shocks capable of producing emission with such properties, 
and thus consider the afterglow radio emission to be strongly influenced 
by scintillation.

Interstellar scintillation effects have been observed in GRB radio light 
curves, and used to obtain indirect measures of the source size 
\citep[for a recent review, see][]{GranotVanDerHorst2014}. 
This method relies on the fact that propagation effects in the interstellar 
medium cause modulations of the flux of a compact source, while a source 
larger than a certain angular size will not vary \citep{Rickett1990}. 
In the case of GRBs, the source is the evolving shock front of the jet, 
which is very compact at first but expands over time. This can result in 
strong modulations at early times, which get quenched at later times 
\citep{Frail1997, Goodman1997, Frail2000}. These variations can be found 
between observations on different days, but intraday variability has also 
been observed in GRBs \citep[e.g.,][]{Chandra2008, VanDerHorst2014}. 
The typical procedure for relating the source size to the scintillation 
effects is to estimate the scintillation strength and timescale using the 
methods of \citet{Walker1998} combined with the NE2001 model of the 
free electrons in our galaxy \citep{CordesLazio2002}. In the 
strong scattering regime, there are two possible types of scintillation: 
refractive and diffractive. In both cases the modulation strength depends 
on the source size compared to the angular scale for scintillation, 
which ranges from a few to a few tens of micro-arcseconds. 
Diffractive scintillation gives stronger flux modulations than refractive 
scintillation, but the angular scale for diffractive scintillation is 
smaller than that for refractive scintillation. Furthermore, the former 
is a narrow-band phenomenon while the latter is broad-band, but they 
could both be at play in GRB afterglow observations.

The redshift of GRB\,151027B is 4.063, which means that 1~arcsecond on the 
sky corresponds to a distance of 7.05~kpc, so 1~micro-arcsecond corresponds 
to $2.2\times10^{16}$~cm. A size of $10^{16}-10^{17}$~cm is typical for the 
jet size, so strong scintillation effects are expected for this GRB, 
also because the high redshift of the GRB means that 40~days in the 
observer frame corresponds to 8~days in the source rest frame. 
The scintillation timescale of several hours to days that we observe for 
GRB\,151027B is plausible, but the observed modulation seems to be 
too large to accommodate within this framework. The maximum modulation 
index for diffractive scintillation is 1, i.e. the flux can increase 
or decrease by a factor of 2 due to scintillation, and the modulation index 
for refractive scintillation is always smaller than 1. Both of those are 
significantly smaller than the jumps in flux that we have observed for 
GRB\,151027B, which are more than a factor of 5 between some observations 
(at the $2\sigma$ level). For instance, at 9~GHz the flux fluctuates 
from $<15\mu$Jy at 14.7~days, to $100\pm10\mu$Jy at 17.7~days, and 
then $<15\mu$Jy at 19.7~days; flux changes of more than a factor of 5, 
both up and down.

Given that these strong flux modulations can not be explained by physical 
processes in the source itself, scintillation does seem to be the most 
natural way to explain the observations, 
as has been done for other GRBs with radio flux modulations. 
However, in this particular case,  we seemingly have to deviate 
from the typical methodology applied in the modeling of scintillation 
effects on GRB radio light curves due to the very large and fast modulations.  
One of the underlying assumptions of 
the usual methodology is that the scattering happens at one location, 
the scattering screen, which resides at a typical distance (usually 1 kpc
from the observer). 
However, many studies of interstellar scintillation with pulsars and 
active galactic nuclei have shown that the distance of the scattering screen 
is quite uncertain. Varying this distance can have strong effects on both
the modulation strength and timescale. For example, some quasars have 
shown extreme intraday variability, 
indicating that their scattering screen is significantly closer than 
what is usually assumed 
\citep{DennettThorpeDeBruyn2002, Bignall2006, MacquartDeBruyn2007, DeBruynMacquart2015}. 
Furthermore, extragalactic sources may be shining through multiple scattering 
screens inside our galaxy, complicating the scintillation behavior even 
further. Every scattering screen will impose its own modulation strength
and timescale, possibly leading to enhanced and complex scintillation 
behaviour.

The bottom line is that the observed fluctuations in GRB\,151027B 
can be explained by scintillation, but the large modulation amplitude
and rapid variations suggests that the scattering screen is at a smaller
distance, that there are multiple screens, or a combination of the two.
It will require many more detailed 
studies of various radio sources, including GRB afterglows, to fully 
probe the scintillation behavior of the interstellar medium in our local
environment.

\begin{figure}[th] 
\includegraphics[width=0.99\columnwidth]{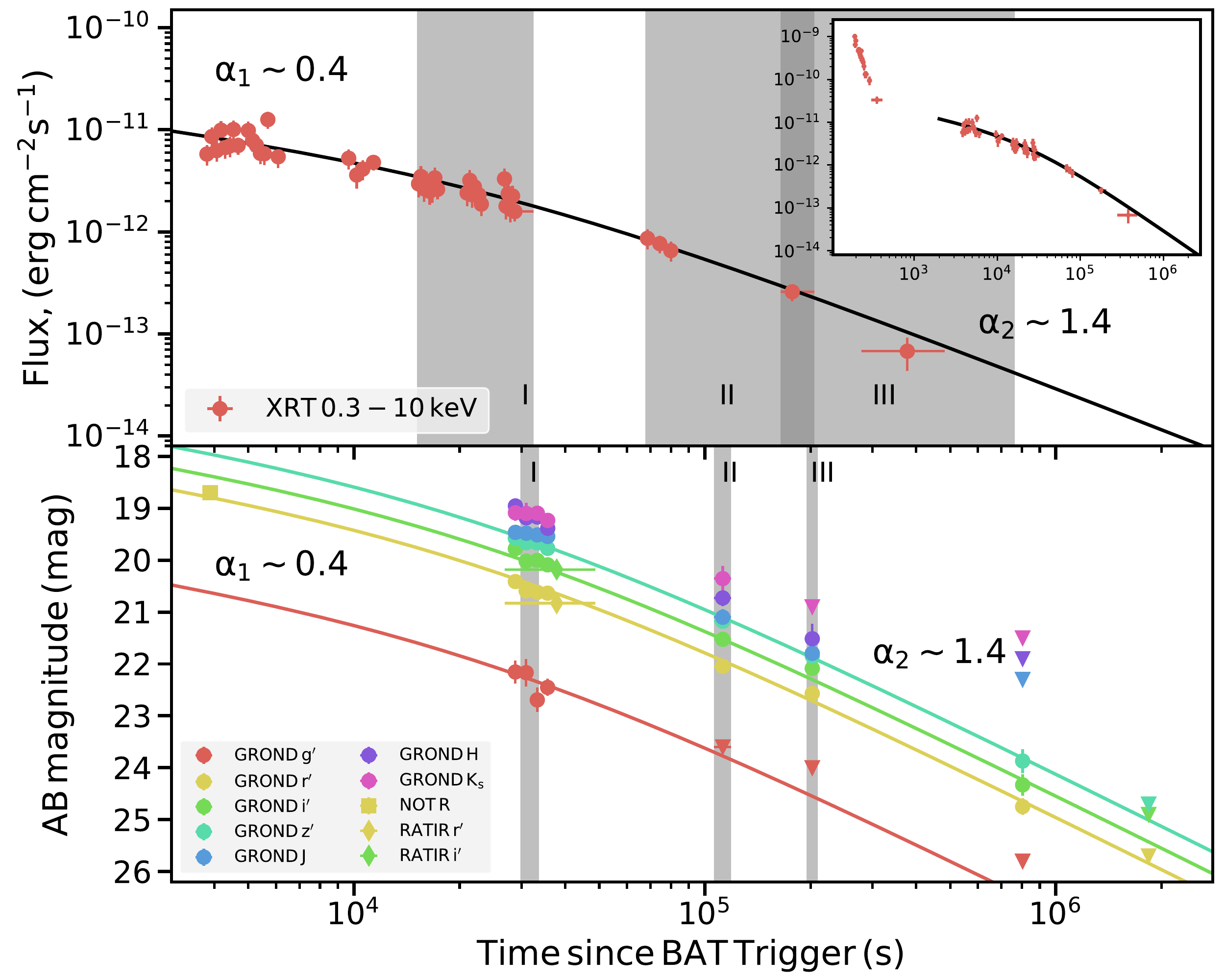} 
\caption[optXlc]{Light curve of the afterglow of GRB 151027B  
at X-rays as observed with Swift/XRT (top), and in the optical/near-infrared
as observed with GROND (no extinction-correction applied), complemented 
with two measurements by NOT and RATIR \citep{Malesani2015, Watson2015}.
Error bars are plotted, but mostly smaller than the symbol size.
The vertical grey bands mark the time intervals for which the
spectral energy distributions have been established 
(see text for more details and Fig. \ref{fig:bbsed}).
\label{fig:optXlc}} 
\end{figure}

\begin{figure}[th]  
\includegraphics[width=0.98\columnwidth]{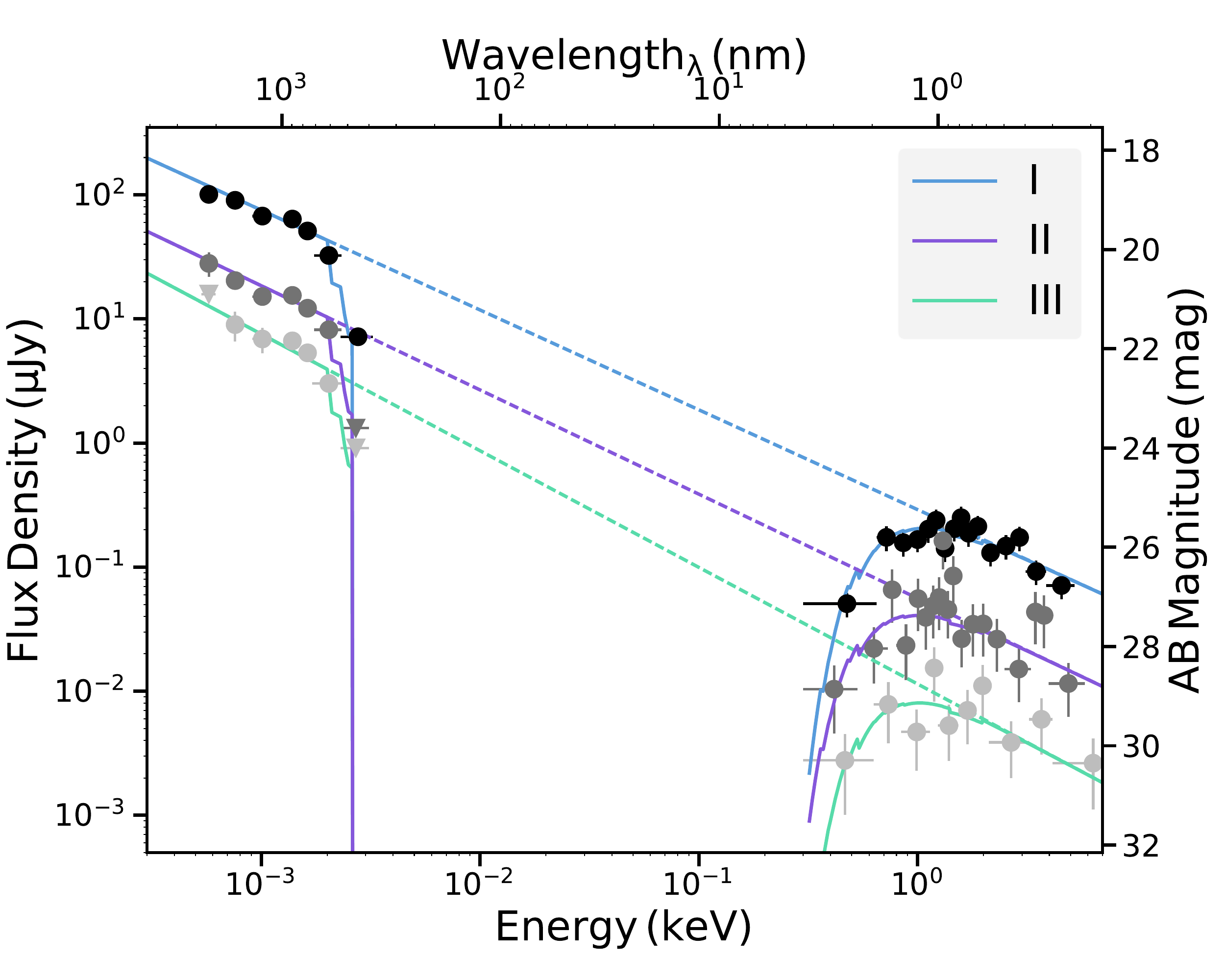}  
\caption[bbsed]{Observer-frame 
optical/near-infrared to X-ray spectral energy distribution 
of the afterglow of GRB 151027B at the three epochs marked in 
Fig. \ref{fig:optXlc} with the grey shading.
Error bars are plotted, but mostly smaller than the symbol size.
\label{fig:bbsed}}  
\end{figure}

\subsection{Constraints on the fireball model}

Both the X-ray and the optical light-curves can be modeled with a smoothly 
broken power-law (Fig. \ref{fig:optXlc}) with
$\alpha_1 = 0.44 \pm 0.19$, $\alpha_2 = 1.44 \pm 0.14$ and $t_b \sim 22.5$ ks, 
consistent with the magnitudes observed by NOT and
RATIR \citep{Malesani2015, Watson2015}. These temporal slopes were used 
to rescale an XRT spectrum from data taken between
$T_0 + 15$ ks and $T_0 + 32$ ks to the stacked GROND data taken between 
$T_0 + 32$ ks and $T_0 + 34$ ks (both these intervals are shaded in grey 
in Fig. \ref{fig:optXlc}). The resulting broadband spectral energy distribution
(SED) is best fit 
with a single power-law of slope $\beta = 0.81 \pm 0.01$, 
with a negligible amount of dust
($A_V = 0.01 \pm 0.01$ mag), independent of the extinction model 
(see \citealt{Bolmer2017} for more details on the extinction determination,
where a broken power-law model has been preferred in order to derive
a conservative extinction value).
In these fits, the \gp\rp\ filters were ignored owing to additional
uncertainty from absorption from the Ly$\alpha$ forest.
The spectral slope in the X-ray to optical/near-infrared does not change
with time within errors 
($\beta_1 = 0.81 \pm 0.01$, $\beta_2 = 0.83 \pm 0.03$ and
$\beta_3 = 0.89 \pm 0.07$)
as evidenced by the other broadband SEDs at later times
(see Fig. \ref{fig:bbsed}), nor do the data
require a spectral break at later times.
The above post-break parameters are fully consistent (within 2$\sigma$)
with an afterglow with 
an electron powerlaw distribution with $p=2.62 \pm 0.02$, evolving 
via slow cooling into an ISM environment where the cooling break is above 
the Swift/XRT upper energy boundary:
measured $\alpha_2 = 1.44 \pm 0.14$ vs. predicted $\alpha_2 = 1.22\pm0.02$.
A cooling break below the GROND bands would imply $p=1.62 \pm 0.02$
and $\alpha_2 = 0.72$, inconsistent with our observed light curve.
In the preferred scenario, the cooling break $\nu_c$ would move to lower 
frequencies
proportional to $t^{-1/2}$. Since we also do not see any signature of a 
spectral break in the X-ray band up to 2$\times$10$^5$ s after the GRB,
after back-extrapolation this implies that $\nu_c$ (31 ks) $>$ 20 {\rm keV}.
We finally note that the pre-break phase is consistent with the 
plateaus seen in many \textit{Swift}-detected GRBs 
\citep[e.g.,][and references therein]{Dainotti+2017},
with the optical data (primarily the NOT data) fitting the picture as well
as evidenced by being in the same synchrotron spectral regime.

The remaining question then is the relative ordering of the peak frequency 
$\nu_m$ and the self-absorption frequency $\nu_{sa}$. 
Given the multiple radio detections
with ATCA implies that the self-absorption frequency should be below 
5 GHz already at 2.8 days after the GRB in order for the scintillation 
amplitude not to exceed a factor of 10. The ALMA limits then require
$\nu_m$ to be above the self-absorption frequency. Considering the
canonical decrease of $\nu_m$ according to $t^{-3/2}$, our following two
observational constraints fix the value of $\nu_m(t)$ to better than 20\%: 
(i) at the time of the first GROND observation, 
$\nu_m$(31 ks)$<1.3\times 10^{14}$ Hz; 
(ii) the ALMA band 7 limit together with the interpolated 
  optical/near-infrared fluxes at this epoch imply 
  $\nu_m$(5.279\,d)$>1.8\times 10^{12}$ Hz.
Back-extrapolating the latter limit to the first GROND observation
(by a factor of (31 ks/456.1 ks)$^{-3/2}$ = 56.4)
implies an inferred $\nu_m$(31 ks)$ = (1.15\pm0.15)\times 10^{14}$ Hz.

With these observational constraints it is possible to determine 
the fireball parameters, as follows: We observe the following set of
relations, all at 31 ks after the GRB:
$$ \nu_m    = (1.15\pm0.15)\times 10^{14} \, {\rm Hz} $$
$$ F_\nu(\nu_m)  = (100\pm 10) \, \mu {\rm Jy}  $$
$$ \nu_c    > 4.8\times 10^{18} \, {\rm Hz} \,\, ( = 20 \, {\rm keV}) $$
$$ F_\nu(\nu_c) < 0.07 \, \mu {\rm Jy} $$

Within the canonical fireball scenario
\citep{grs02} in the slow cooling case with the ordering
$\nu_{sa} < \nu_m < \nu_c$
and ISM density profile, the self-absorption frequency remains constant, 
and being always below 100 MHz, i.e. below our observed frequencies,
for all the allowed parameter range (see below), it does not provide any 
further constraint.

\begin{figure*}[th]
\hspace{-0.9cm} 
\vspace{-0.3cm}
\includegraphics[width=0.77\textwidth, angle=270]{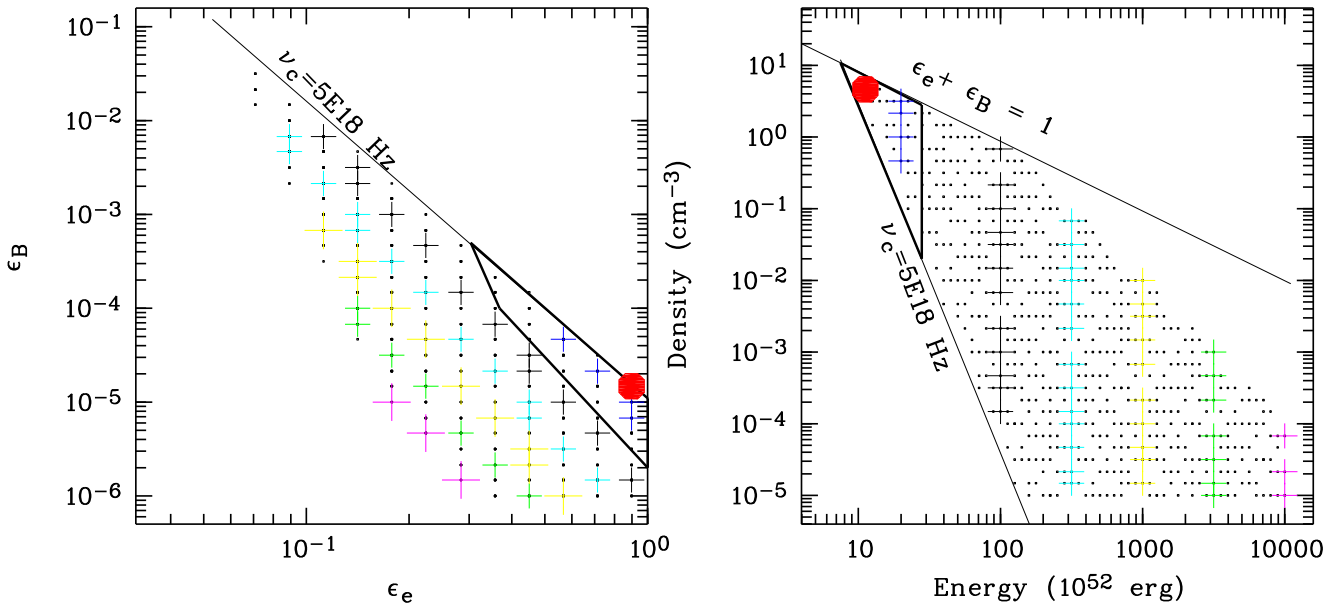} 
\vspace{-4.5cm} 
\caption[151027Bepseps]{Constraints on the fireball parameters for the 
afterglow of GRB 151027B. Small black dots delineate the allowed phase space
by the four constraints at 31 ks,
and colored crosses visualize different possible solutions for seven
different values of the kinetic energy. The solution with the lowest
kinetic energy is marked with the red-filled octagon: 
$E_{\rm kin} = 11.2 \times 10^{52}$ erg, external density $n = 5$ cm$^{-3}$,
$\epsilon_e = 0.9$ and
$\epsilon_B = 1.5 \times 10^{-5}$. 
The thick-lined triangles enclose the allowed parameter range,
if $E_{\rm kin} < 5 \times E_{\rm \gamma, iso}$.
}
\label{fig:param}
\end{figure*}

These constraints on the observed frequencies and fluxes lead to 
bounds on the fireball parameters as visualized in Fig. \ref{fig:param}.
While the observations do not uniquely constrain all parameters
we can use an efficiency argument to derive a likely parameter range.
In the standard picture, a fraction $\epsilon_\gamma$ of the explosion energy
is radiated in the prompt radiation (observable as $E_{\rm \gamma, iso}$),
and the remaining fraction ending up as kinetic energy $E_{\rm kin}$
of the swept up ambient gas. Early observations suggested nearly
equipartition between these two channels, though later considerations
including proper error estimates suggest $\epsilon_\gamma$ in the range
of $\sim 0.1-0.5$ \citep{Granot+2006}. Assuming $\epsilon_\gamma = 0.2$ and 
using $E_{\rm \gamma, iso}$ (15--10000 keV) = $(5 \pm 1) \times 10^{52}$ erg 
(based on a best-fit cut-off powerlaw fit of the Swift/BAT data provided at
http://gcn.gsfc.nasa.gov/notices\_s/661869/BA/ giving
an energy fluence of (14.7$\pm$2.6) $\times 10^{-7}$ erg cm$^{-2}$),
the GRB 151027B fireball parameters are constrained as follows:
external density $n = 0.03-5$ cm$^{-3}$, 
$\epsilon_e  = 0.3-1.0$ and  
$\epsilon_B = 4 \times 10^{-4} - 2 \times 10^{-6}$ 
(see Fig. \ref{fig:param}).
We note that our derived $\epsilon_e$ is higher than the majority of
published afterglows, though still in the allowed range.

\begin{figure}[th]
\includegraphics[width=1.01\columnwidth]{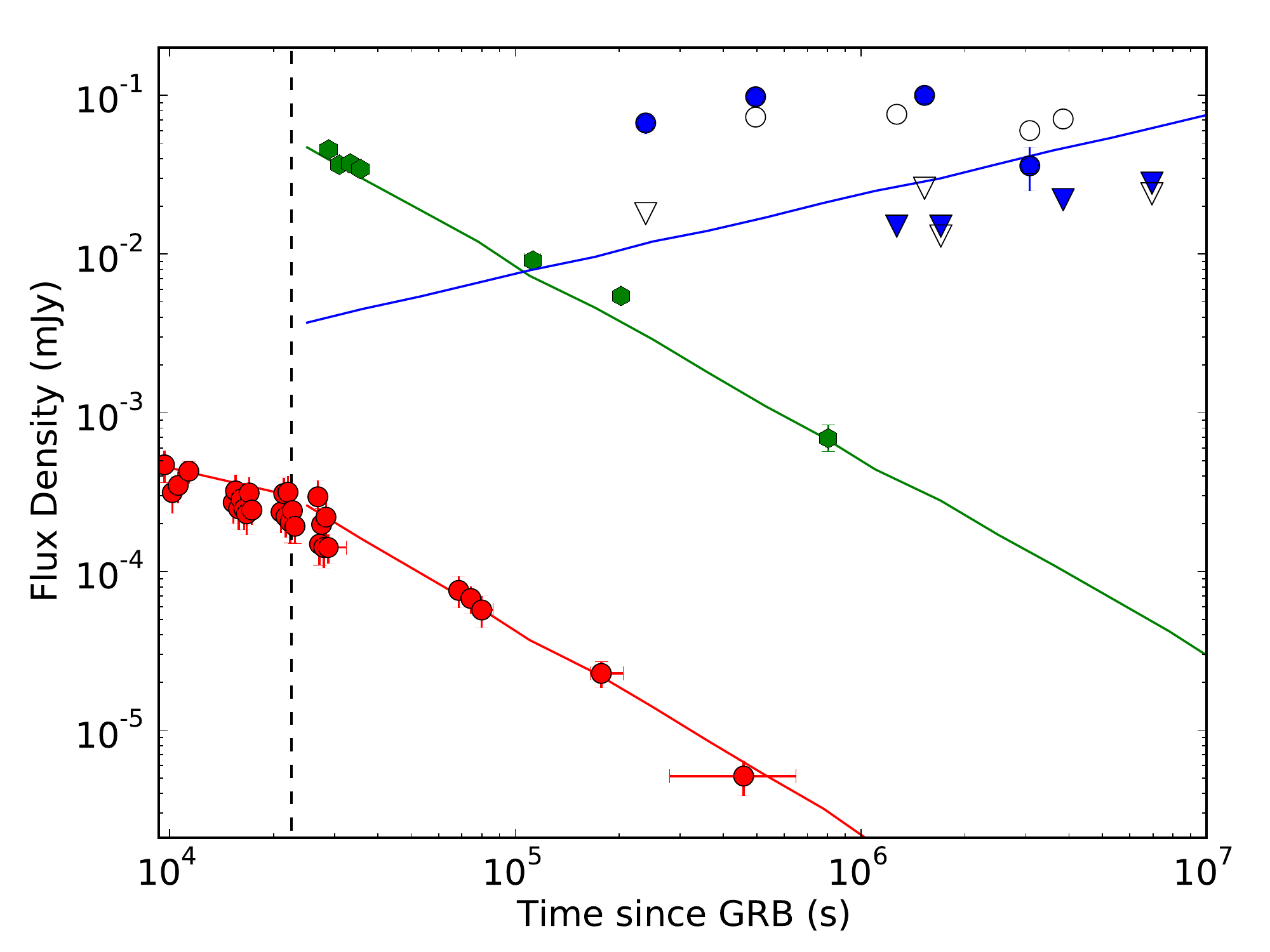} 
\caption[151027Bmodradio]{Model light curves for the
afterglow of GRB 151027B at X-rays (red), optical (green, extinction-corrected)
and radio (blue; open/filled symbols are 5.5/9 GHz, respectively; 
circles=detection, triangles=upper limits)
for the lowest-$E_{\rm kin}$ parameter set. Data are drawn with error bars, 
but those are mostly smaller than the symbol size.
The vertical line denotes the break time of 22.5 ks (see sect. 3.2).
Radio data have not been used
in deriving the model, so the blue curve is actually a `predicted' 
light curve. The fact that it matches the mean between the radio 
measurements and upper limits demonstrates that the model is 
reasonably good, and that the scintillation interpretation for the
radio measurements is reasonable.
}
\label{fig:modradio}
\end{figure}

For the solution with the lowest kinetic energy, 
$E_{\rm kin} = 11.2 \times 10^{52}$ erg (remaining parameters see caption of
Fig. \ref{fig:param}), we then compute the X-ray, optical I-band and 
7 GHz radio (as average  between 5.5 and 9 GHz) light curves. These are 
shown in Fig. \ref{fig:modradio}. Note that the model was derived without
using constraints from the radio bands, so it is interesting that the
'predicted' radio light curve of Fig. \ref{fig:modradio} corresponds 
roughly to the mean of the radio detections and upper limits.
This implies that the scintillation
interpretation of the measured radio fluxes is reasonable.
Thus,  the observed scintillation corresponds to about a factor 
$\pm$3 variation in either direction, but is not a one-way excursion.

\section{Conclusions}

Our data set of the GRB 151027B afterglow can be explained with
the simplest version of the standard fireball scenario, displaying a 
single synchrotron spectrum evolving according to standard dynamics. 

We find that the blast wave moves into
a constant-density environment, in the slow cooling regime.
While we do not see the characteristic movement of the cooling frequency
in or through the X-ray band, the constancy of the peak flux over
the full observing epoch as evidenced by our data requires a constant
density profile. The derived fireball parameters are all within the
range expected and discussed in the literature.
For the smallest allowed kinetic energy, $\epsilon_e$ is pushed towards
the upper limit of 1.

After GRBs 000131, 050904, 090423, 111008A, 120521C, 130606A, 140304A, 
140311A, 140515A the afterglow of GRB 151027B 
is the tenth above a redshift of 4 which is detected in the radio band
(see the summary table http://www.mpe.mpg.de/$^\sim$jcg/grbgen.html).
Its peak spectral radio luminosity (2$\times$10$^{31}$ erg/s/Hz) 
is among the top quarter of radio afterglows \citep{ChandraFrail2012},
but certainly not exceptional. However, the large-amplitude
and rapid flux fluctuations up to 9 GHz are exceptional, and
imply that scintillation plays 
a major role, even at 45 days (9 days rest-frame) post-burst.

We finally mention that the ALMA flux limits are close to the prediction
of our model, so we cannot completely rule out that the 3.5$\sigma$ 
blob in the band 7 image is not, indeed, the afterglow. 

With the above caveats it is worth noting that this is one of the few
afterglows of long-duration GRBs for which the simplest version of 
the afterglow scenario describes a rather extensive multi-epoch and
multi-frequency data set. In many cases, more data also meant a 
need for a more complicated afterglow scenario. This is independent of
the publication bias that afterglows with 'exciting' irregular behaviour
(like GRBs 071031 -- \citealt{kgm09}, 080129 -- \citealt{gkm09}, 
081029 -- \citealt{Nardini2011}, 100621A -- \citealt{gkn13}, 
100814A -- \citealt{Nardini2014}, 111209A -- \citealt{gmk15, Kann+17})
get more easily published than standard GRB afterglows.
It remains to be investigated whether or not some of the standard
afterglows can be fitted with the next-simplest version of afterglow
models. 
The hydrodynamical simulations including the  
incorporation of the off-axis angle view \citep{vanEerten2015}
represent one way, analytical jet spreading models another. 

While there is a wealth of published papers dealing with the
(fireball) modelling of individual GRB afterglows, the vast majority only
allows consistency checks, since the data are not sufficient to
derive all five (plus redshift) model parameters. The three
historical exceptions for which all parameters could be determined 
are GRBs 980703 \citep{Frail2003}, 000926 \citep{Harrison2001} and
090323 \citep{Cenko11}. More recently, our group managed to add
another 4 GRBs to this sample (100418A, 110715A, 121024A, 130418A --
\citep{Varela2017}; for GRB 121024A see \cite{Varela+2016} for details.
This small sample, out of a total of $>$700 known 
X-ray/optical afterglows, shows the challenge of testing the
afterglow model(s). And even those seven GRB afterglows are not
uniquely described by a single set of parameters or the simplest
fireball version: one GRB is equally well described by either 
wind or ISM density profile, two other GRBs show substantial
flaring activity implying additional energy injection, and another
two GRBs show strong evidence for an inverse Compton component.
There are indications in this sample for a preference of a 
wind-like GRB environment, contrary to the results of many early analyses.
However, this topic as well as other afterglow-related issues like
a reliable distribution of 
microphysical parameters, require a substantially larger sample size.

From an observer point of view, it is not obvious
how to best reach a larger sample size, i.e. how a guaranteed-success 
strategy would look like.
Radio observations are only meaningful at late times when scintillation 
has ceased, but then the X-ray and optical/NIR instrumentation typically
is not sensitive enough to detect the afterglow anymore. Yet, radio
observations provide crucial constraints for the afterglow modelling.
Alternatively, dedicated multi-band, multi-epoch ALMA monitoring seems 
promising for two reasons:
Firstly, it is sensitive enough to cover a larger time interval of the
afterglow emission (say 2-3 weeks).
Secondly, the $\nu_m$-crossing is faster than that of $\nu_c$, allowing
(in combination with the decay slope) a potentially better distinction 
between wind and ISM environment. 
While rapid (within a day) 
target-of-opportunity (ToO) observations are finally allowed with ALMA, 
the general acceptance level of GRB-related ToO-proposals is going down 
after more than a decade of {\it Swift}-driven afterglow studies, and the 
need for getting proposals accepted at several observatories during the 
same semester does not make things easier (see e.g. 
\citealt{Middleton2017} for a description of the problem and suggested
solutions).
Instead of attempting full multi-wavelength coverage over a long 
time-interval, a graded approach with dense X-ray/optical/sub-mm
coverage during the first days and sub-mm/radio at late stages
might be the better approach. This is particularly motivated by the
potential of 
trans-relativistic dynamical models and models including jet dynamics,
that improve upon earlier closure relations.
The number of open questions and the impact which a proper knowledge
of the GRB afterglow emission process would have on a variety of
other astrophysical areas certainly justifies a concerted approach.

\begin{acknowledgements}
During the writing of this paper, the
radio astronomy community lost a great scientist, and we lost a dear  
colleague and collaborator, Ger de Bruyn. His insights in radio
scintillation were crucial in the work presented in this paper, and he
will be missed by many.

JG is particularly grateful to Phil Edwards for scheduling the many 
ATCA ToO observations.

SK, ANG, SS and DAK acknowledge support by DFG grant Kl 766/16-1.
DAK acknowledges financial support from the Spanish research project
AYA 2014-58381-P, and from Juan de la Cierva Incorporaci\'on fellowships 
IJCI-2015-26153 and IJCI-2014-21669.
JFG, TK and PW acknowledges support through the Sofja Kovalevskaja award to 
P. Schady from the A. von Humboldt foundation of Germany.

Part of the funding for GROND (both hardware as well as personnel)
was generously granted from the Leibniz-Prize to Prof. G. Hasinger
(DFG grant HA 1850/28-1).

The Australia Telescope Compact Array is part of the Australia Telescope 
National Facility which is funded by the Commonwealth of Australia for 
operation as a National Facility managed by CSIRO.

ALMA is a partnership of ESO (representing its member states), NSF (USA) 
and NINS (Japan), together with NRC (Canada) and NSC and ASIAA (Taiwan) 
and KASI (Republic of Korea), in cooperation with the Republic of Chile. 
The Joint ALMA Observatory is operated by ESO, AUI/NRAO and NAOJ.

This work made use of data supplied by the UK \textit{Swift} Science Data
Centre at the University of Leicester.

\end{acknowledgements}

\noindent {\small {\it Facilities:} Max Planck:2.2m (GROND), ATCA,
                  ALMA, Swift }

\begin{appendix}

\begin{table*}[th]
   \caption{Secondary standard stars used to calibrate the optical/NIR 
      afterglow measurements.
      The \gp\rp\ip\zp\ magnitudes are in the AB system, $JHK_{\rm s}$ in Vega.}
   \vspace{-0.2cm}
   \tiny
      \begin{tabular}{@{\kern5pt}c@{\kern8pt}c@{\kern8pt}c@{\kern5pt}c@{\kern5pt}c@{\kern5pt}c@{\kern5pt}c@{\kern5pt}c@{\kern5pt}c}
      \hline
      \noalign{\smallskip}
      RA/Decl. (deg) & RA/Decl. (hms) & \gp & \rp & \ip & \zp &   $J$ &  $H$  & $K_{\rm s}$ \\ 
      (2000.0)       & (2000.0)       &(mag)&(mag)&(mag)&(mag)&(mag)&(mag)&(mag)  \\
      \noalign{\smallskip}
      \hline
      \noalign{\smallskip}
$\!\!\!$76.22154 -6.41494 & 05:04:53.17 -06:24:53.8 & 16.838$\pm$0.001 & 15.947$\pm$0.001 & 15.664$\pm$0.001 & 15.440$\pm$0.001 & 14.437$\pm$0.002 & 14.000$\pm$0.002 & 13.860$\pm$0.009\\
$\!\!\!$76.23341 -6.42239 & 05:04:56.02 -06:25:20.6 & 15.289$\pm$0.001 & 14.961$\pm$0.001 & 14.475$\pm$0.001 & 14.325$\pm$0.001 & 13.375$\pm$0.001 & 13.022$\pm$0.002 & 12.860$\pm$0.004\\
$\!\!\!$76.18746 -6.43186 & 05:04:44.99 -06:25:54.7 & 15.155$\pm$0.001 & 14.914$\pm$0.001 & 14.441$\pm$0.001 & 14.308$\pm$0.001 & 13.374$\pm$0.001 & 13.023$\pm$0.002 & 13.047$\pm$0.004\\
$\!\!\!$76.18821 -6.43417 & 05:04:45.17 -06:26:03.0 & 18.943$\pm$0.009 & 18.278$\pm$0.004 & 18.046$\pm$0.005 & 17.858$\pm$0.005 & 16.910$\pm$0.020 & 16.439$\pm$0.027 & 16.338$\pm$0.089\\
$\!\!\!$76.22938 -6.43628 & 05:04:55.05 -06:26:10.6 & 16.412$\pm$0.001 & 15.793$\pm$0.001 & 15.512$\pm$0.001 & 15.323$\pm$0.001 & 14.381$\pm$0.002 & 13.936$\pm$0.002 & 13.853$\pm$0.009\\
$\!\!\!$76.20600 -6.44189 & 05:04:49.44 -06:26:30.8 & 19.557$\pm$0.014 & 17.987$\pm$0.003 & 17.294$\pm$0.003 & 16.894$\pm$0.002 & 15.656$\pm$0.005 & 14.906$\pm$0.007 & 14.848$\pm$0.023\\
$\!\!\!$76.20441 -6.44311 & 05:04:49.06 -06:26:35.2 & 18.429$\pm$0.005 & 16.989$\pm$0.001 & 16.419$\pm$0.001 & 16.063$\pm$0.001 & 14.871$\pm$0.004 & 14.159$\pm$0.004 & 14.115$\pm$0.012\\
$\!\!\!$76.24538 -6.44383 & 05:04:58.89 -06:26:37.8 & 18.927$\pm$0.007 & 17.926$\pm$0.003 & 17.539$\pm$0.003 & 17.282$\pm$0.003 & 16.177$\pm$0.007 & 15.552$\pm$0.009 & 15.483$\pm$0.037\\
$\!\!\!$76.24888 -6.44475 & 05:04:59.73 -06:26:41.1 & 18.823$\pm$0.007 & 17.953$\pm$0.003 & 17.592$\pm$0.003 & 17.350$\pm$0.003 & 16.287$\pm$0.009 & 15.775$\pm$0.012 & 15.698$\pm$0.046\\
$\!\!\!$76.24046 -6.44611 & 05:04:57.71 -06:26:46.0 & 18.042$\pm$0.003 & 17.457$\pm$0.002 & 17.242$\pm$0.002 & 17.068$\pm$0.002 & 16.179$\pm$0.007 & 15.785$\pm$0.012 & 15.733$\pm$0.046\\
$\!\!\!$76.24809 -6.44664 & 05:04:59.54 -06:26:47.9 & 18.253$\pm$0.004 & 17.550$\pm$0.002 & 17.278$\pm$0.002 & 17.084$\pm$0.003 & 16.097$\pm$0.007 & 15.634$\pm$0.011 & 15.586$\pm$0.039\\
$\!\!\!$76.24358 -6.45283 & 05:04:58.46 -06:27:10.2 & 16.975$\pm$0.001 & 16.058$\pm$0.001 & 15.684$\pm$0.001 & 15.422$\pm$0.001 & 14.353$\pm$0.002 & 13.896$\pm$0.002 & 13.711$\pm$0.007\\
$\!\!\!$76.20421 -6.45503 & 05:04:49.01 -06:27:18.1 & 16.637$\pm$0.001 & 16.080$\pm$0.001 & 15.909$\pm$0.001 & 15.760$\pm$0.001 & 14.888$\pm$0.002 & 14.595$\pm$0.005 & 14.589$\pm$0.014\\
$\!\!\!$76.24854 -6.45628 & 05:04:59.65 -06:27:22.6 & 17.206$\pm$0.002 & 16.637$\pm$0.001 & 16.419$\pm$0.001 & 16.259$\pm$0.001 & 15.334$\pm$0.004 & 15.057$\pm$0.007 & 14.938$\pm$0.018\\
$\!\!\!$76.24492 -6.45639 & 05:04:58.78 -06:27:23.0 & 19.816$\pm$0.017 & 18.618$\pm$0.005 & 18.125$\pm$0.005 & 17.828$\pm$0.005 & 16.660$\pm$0.014 & 16.090$\pm$0.018 & --\\
$\!\!\!$76.20492 -6.45661 & 05:04:49.18 -06:27:23.8 & 17.422$\pm$0.002 & 16.985$\pm$0.001 & 16.840$\pm$0.002 & 16.710$\pm$0.002 & 15.832$\pm$0.007 & 15.538$\pm$0.011 & 15.668$\pm$0.039\\
$\!\!\!$76.23579 -6.46542 & 05:04:56.59 -06:27:55.5 & 17.697$\pm$0.002 & 16.485$\pm$0.001 & 16.035$\pm$0.001 & 15.720$\pm$0.001 & 14.555$\pm$0.002 & 13.993$\pm$0.002 & 13.858$\pm$0.007\\
$\!\!\!$76.20871 -6.46989 & 05:04:50.09 -06:28:11.6 & 17.155$\pm$0.002 & 15.646$\pm$0.001 & 14.865$\pm$0.001 & 14.477$\pm$0.001 & 13.231$\pm$0.001 & 12.624$\pm$0.001 & 12.455$\pm$0.002\\
$\!\!\!$76.22787 -6.47075 & 05:04:54.69 -06:28:14.7 & 19.456$\pm$0.014 & 17.722$\pm$0.002 & 16.879$\pm$0.002 & 16.392$\pm$0.001 & 15.037$\pm$0.004 & 14.370$\pm$0.004 & 14.142$\pm$0.009\\
$\!\!\!$76.25004 -6.47411 & 05:05:00.01 -06:28:26.8 & 14.953$\pm$0.001 & 14.683$\pm$0.001 & 14.181$\pm$0.001 & 13.959$\pm$0.001 & 12.881$\pm$0.001 & 12.503$\pm$0.001 & 12.362$\pm$0.002\\
$\!\!\!$76.21046 -6.48039 & 05:04:50.51 -06:28:49.4 & 20.958$\pm$0.052 & 19.300$\pm$0.010 & 18.445$\pm$0.007 & 17.950$\pm$0.005 & 16.676$\pm$0.012 & 15.973$\pm$0.014 & 15.843$\pm$0.043\\
$\!\!\!$76.20200 -6.48617 & 05:04:48.48 -06:29:10.2 & 20.425$\pm$0.035 & 18.816$\pm$0.007 & 17.941$\pm$0.005 & 17.449$\pm$0.004 & 16.118$\pm$0.007 & 15.430$\pm$0.009 & 15.310$\pm$0.027\\
     \noalign{\smallskip}
      \hline
   \end{tabular}
   \label{Stdstars}
\end{table*}

\end{appendix}

\end{document}